\documentclass[reprint, amsmath,amssymb,aps,prx,floats,floatfix]{revtex4-1}

\usepackage{graphicx}
\usepackage{bm}
\usepackage{bbm}
\usepackage{xcolor}

\newcommand{\comm}[2]{[#1,#2]}
\newcommand{\acomm}[2]{\{#1,#2\}}
\newcommand{\ket}[1]{\left\lvert{#1}\right\rangle}
\newcommand{\bra}[1]{\left\langle{#1}\right\rvert}
\newcommand{\dyad}[1]{\ket{#1}\bra{#1}}

\newcommand{\expval}[1]{\left\langle{#1}\right\rangle}

\newcommand{\Tr}[1]{\mathrm{tr}\,#1}

\newcommand{\revision}[1]{#1}

\begin{document}

\title{Many-body localization of zero modes}

\author{Christian P. Chen}
\affiliation{Department of Physics, Lancaster University, Lancaster, LA1 4YB, United Kingdom}
\author{Marcin Szyniszewski}
\affiliation{Department of Physics, Lancaster University, Lancaster, LA1 4YB, United Kingdom}
\author{Henning Schomerus}
\affiliation{Department of Physics, Lancaster University, Lancaster, LA1 4YB, United Kingdom}

\date{\today}
\begin{abstract}
The celebrated Dyson singularity signals the relative delocalization of single-particle wave functions at the zero-energy symmetry point of disordered systems with a chiral symmetry. Here we show that analogous zero modes in interacting quantum systems can \revision{fully} localize at sufficiently large disorder, but do so less strongly than nonzero modes, as signified
by their \revision{real-space} and Fock-space localization characteristics.
We demonstrate this effect in a spin-1 Ising chain, which naturally provides a chiral symmetry in an odd-dimensional Hilbert space, thereby guaranteeing the existence of a many-body zero mode at all disorder strengths.
\revision{In the localized phase, the bipartite entanglement entropy of the zero mode follows an area law, but is enhanced by a system-size-independent factor of order unity when compared to the nonzero modes.}
\revision{Analytically, this feature can be }attributed to a specific zero-mode hybridization pattern on neighboring spins.
The zero mode also displays a symmetry-induced even-odd and spin-orientation
fragmentation of excitations, characterized by real-space spin correlation functions, which
generalizes the sublattice polarization of topological zero modes in noninteracting systems, and
holds at any disorder strength.
\end{abstract}

\maketitle

\section{Introduction}
Complex quantum systems owe their rich physical properties to the intricate interplay of symmetries, disorder, and interactions.
This  interplay is mirrored by the intertwined concepts and frameworks which have been developed to capture these aspects.
Symmetry-reduced representations of quantum systems were established at the beginning of quantum mechanics
\cite{weyl1946classical},
whilst the absence
of unitary symmetries allows for complex  wave dynamics even for low numbers of degrees of freedom
\cite{Haake2019}. This ties both to
semiclassical descriptions of classically chaotic systems as well as to statistical descriptions of structureless
noninteracting disordered systems, whose properties are captured by random-matrix theory \cite{Mehta2004}.
The latter provides a natural framework to classify
complex quantum systems also in accordance with their invariance under
antiunitary symmetries, as pioneered by Wigner and Dyson \cite{Wigner1955,dyson1962threefold} and completed
with the ten Altland-Zirnbauer universality classes \cite{PhysRevB.55.1142}. This ten-fold way also underpins the topological
classification of electronic band structures in periodic systems of different spatial dimensionality
\cite{RevModPhys.88.035005}. Besides time-reversal symmetry, this classification also accounts for antiunitary charge-conjugation symmetries and the combination of the two into a unitary chiral symmetry, as originally encountered in random-matrix descriptions of Dirac systems \cite{PhysRevLett.70.3852}.

For structureless systems,
random-matrix theory describes wave
functions ergodically spreading across the whole system, but this can be amended to also provide statistical descriptions of Anderson-localized
systems, in particular in one-dimensional or quasi-one-dimensional geometries \cite{RevModPhys.69.731}.
Again, these descriptions can be organized according to the universality classes of the ten-fold way
\cite{PhysRevB.63.235318}, and then account for topological phenomena in nonperiodic, disordered settings. The most striking effect amongst these is the possibility of such systems to \revision{be less localized} near spectral symmetry points. This phenomenon was first realized by Dyson \cite{Dyson1953}, who noted that a one-dimensional system with hopping disorder develops a logarithmically diverging density of states around the band center---the so-called Dyson singularity, which goes along with \revision{anomalously localized states that exhibit a stretched-exponential spatial profile.}
Within the classification framework described above, this \revision{relative} delocalization phenomenon becomes tied to the existence of a topologically protected zero mode in a chirally symmetric system with an odd-dimensional Hilbert space \cite{PhysRevLett.81.862}\revision{, and also occurs in higher-dimensional systems, where the anomalously localized states can resemble those at a metal-insulator transition \cite{Hatsugai1997,Xiong2001}.}
Weaker analogues of such \revision{anomalous localization} characteristics also occur in absence of spectral symmetries \cite{kappus1981anomaly,PhysRevB.67.100201}. Such robust \revision{features} deserve attention as they significantly broaden the scope for topologically protected quantum phenomena to realistic, disordered systems, both conceptually as well as in practical terms.

For many-body systems, interactions provide a significant complication of all of these aspects, with much recent effort devoted to the question of ergodic versus many-body localized behavior \cite{Altman2015,Nandkishore2015,Abanin2017}. The ergodic phase is again well captured by random-matrix theory \cite{Page1993}, reflecting its original setting of nuclear physics \cite{Wigner1956}. Topological aspects have been extensively studied
for gapped ground-state physics \cite{wen2004quantum}, but topological order can also emerge in excited states, where it competes with the many-body localized phase \cite{PhysRevB.88.014206,PhysRevLett.113.107204}. In particular, it has been found that such novel phases can be induced by a  particle-hole symmetry that pairs up excited states around the centre of the spectrum \cite{PhysRevLett.113.107204,vasseur2016,tomasi2020}. However, the role of many-body zero modes pinned to such spectral symmetry points is much less understood, both conceptually as well as concerning the identification of concrete phenomena.

Here, we clarify this role by drawing motivation from  the Dyson singularity.
We first identify a natural disordered many-body system displaying an analogous chiral many-body zero mode, consisting of a simple spin-1 Ising chain with a random transverse field. We then address the question of its localization properties both in real space and in Fock space, where we identify two localization phenomena that characterize the zero mode. (i) In real space the spin correlations of the zero mode fragment into 5 independent sectors. This fragmentation occurs both with respect to the even and odd sublattice, which generalizes the sublattice polarization of chiral zero modes in noninteracting systems, as well as with respect to  the  orientations of the correlated spin, which is specific to the chosen many-body context. This phenomenon holds at all disorder strengths.
(ii) In contrast to the noninteracting case, the zero mode still localizes at strong disorder\revision{, both in real space as well as in Fock space, as indicated, e.g., by an area law for the entanglement entropy, a large inverse participation ratio, and  short-ranged spatial correlations.} \revision{However, these measures also indicate that quantitatively, the zero mode is noticeably less localized than the nonzero modes---a phenomenon that for brevity we refer to as `relative delocalization' in the remainder of this work. In particular,  the bipartite entanglement entropy is significantly enhanced for the zero mode by a system-size-independent factor of order unity, whilst the inverse participation ratio is correspondingly reduced.}
Thereby, the zero mode attains characteristics that set it apart from all other states in the system, even if they may be very close in energy.

In Sec.~\ref{sec:background}, we describe the spin-1 Ising chain, which provides a natural model for the described phenomena as it combines a chiral symmetry with an odd-dimensional Hilbert space, and always features a zero mode in one of the two spin-parity sectors.
In Sec.~\ref{sec:fragmentation} we first discuss the real-space fragmentation of the spin correlations, as this follows directly from the symmetry constraints and holds at all disorder strengths, which we show analytically and illustrate numerically. In this section we also identify the spin correlations that are most characteristic to quantify the localization properties of zero modes and nonzero modes, to which we then turn to in Secs.~\ref{sec:delocalization}  and \ref{sec:pert}.
In Sec.~\ref{sec:delocalization} we demonstrate the relative delocalization of the zero mode numerically based on both Fock-space and real-space measures. The analytical explanation of this relative delocalization is provided in Sec.~\ref{sec:pert}, where we identify the dominant hybridization patterns of zero modes and nonzero modes. Enforced by the chiral-symmetry constraints, the dominant zero-mode hybridization configurations involve three spin states on neighboring spins, whilst those of nonzero modes only involve two spin states, so that the Fock-space localization characteristics of these modes fundamentally differ. In Sec.~\ref{sec:conclusions} we summarize and discuss the results and put them into further context.

\section{Background}
\label{sec:background}
\subsection{Model}

To demonstrate the effects outlined in this work, we require a many-body system in which a chiral symmetry is manifest for a system with an odd-dimensional Hilbert space. This is naturally provided by a spin-1 Ising chain with a transverse magnetic field, given by the Hamiltonian
\begin{equation}
H = \sum_{n=1}^{N}h_{n}S_{n}^{z} + J\sum_{n=1}^{N}S_{n}^{x}S_{n+1}^{x}.
\label{eqn:Ising_1}
\end{equation}
Here $J$ is the coupling strength between adjacent spins, described by the matrices
\begin{align}
	& S^{x} = \frac{1}{\sqrt{2}}
	\begin{pmatrix}
		0 & 1 & 0 \\
		1 & 0 & 1 \\
		0 & 1 & 0
	\end{pmatrix},
	& S^{y} = \frac{1}{\sqrt{2}i}
	\begin{pmatrix}
		0 & 1 & 0 \\
		-1 & 0 & 1 \\
		0 & -1 & 0
	\end{pmatrix},
\nonumber \\
	& S^{z} =
	\begin{pmatrix}
		1 & 0 & 0 \\
		0 & 0 & 0 \\
		0 & 0 & -1
	\end{pmatrix},
\label{eqn:S1mat_1}
\end{align}
while disorder of strength $W$ is introduced via the on-site potentials, chosen independently from uniform box distributions $h_{n}\in [-W,W]$.
The length of the chain is denoted as $N$, and in our general discussion below the size of the individual spins is denoted as $S$. The Hilbert space then has dimension $\mathcal{N}=(2S+1)^{N}$, and is conveniently spanned by the joint eigenbasis
\begin{equation}
|\mathbf{s}\rangle=\bigotimes_n|s^z_n\rangle
\label{eq:basis}
\end{equation}
 of all operators $S_n^z$, where we label the states by the vector $\mathbf{s}$ of components $s^z_n\in\{-S,-S+1,\ldots,S\}$.

\subsection{Symmetries}\label{sec:Chiral_1}

The primary reason we choose to study the transverse Ising model owes to the fact that it possesses a chiral symmetry
\begin{equation}
	\mathcal{X}H\mathcal{X} = -H,
\label{eqn:Chiral_1}
\end{equation}
with a unitary involution fulfilling $\mathcal{X}\mathcal{X}^\dagger = \mathcal{X}^{2} = \openone$.
To see that this chiral symmetry is manifest for all sizes of spin, consider the spin rotation operator
\begin{equation}
\mathcal{X}=i^{2SN} \prod_{n~\mathrm{even}} U_n^{x}(\pi )\prod_{n~\mathrm{odd}}U_n^{y}(\pi ),
\label{eqn:GenU_1}
\end{equation}
with individual rotation matrices
\begin{equation}
U_n^{a}(\varphi ) = \exp\left(i\varphi S_n^{a}\right).
\label{eqn:genxyz_1}
\end{equation}
The operator~\eqref{eqn:GenU_1} is well-defined in infinite chains and finite chains with open boundary conditions, while periodic boundary conditions require that the chain consists of an even number of spins, as we will thus assume.
The operator $\mathcal{X}$ rotates all spins by $\pi$ about axes  alternatingly aligned with $x$ and $y$, which inverts the sign of all on-site terms $\sim S_n^z$, as well as exactly one of the two spin operators in the interaction terms $\sim S_n^xS_{n+1}^x$, in accordance with the requirements of Eq.~\eqref{eqn:Chiral_1}.
The factor $i^{2SN}$ in the definition \eqref{eqn:GenU_1} makes sure that $\mathcal{X}^2=\openone$ holds for all system lengths and arbitrary spin.

The assignment of sites as even or odd provides a gauge freedom, allowing the definition of an alternative chiral symmetry operator
\begin{equation}
\tilde{\mathcal{X}}=i^{2SN}\prod_{n~\mathrm{even}} U_n^{y}(\pi )\prod_{n~\mathrm{odd}}U_n^{x}(\pi ).
\label{eqn:GenU_1a}
\end{equation}
It follows that the system also possesses an ordinary symmetry $[\mathcal{P},H]=0$, arising from
\begin{equation}
\mathcal{X}\tilde{\mathcal{X}}\propto \mathcal{P}\equiv\prod_nU_n^z(\pi),
\end{equation}
which inverts the sign of all operators $S^x_n$ in the Hamiltonian and leaves the operators $S^z_n$ invariant.
In the basis \eqref{eq:basis},
\begin{equation}
\mathcal{P}|\mathbf{s}\rangle=\exp(i\pi \sum_n s_n^z)|\mathbf{s}\rangle=(-1)^{\sum_n s_n^z}|\mathbf{s}\rangle,
\end{equation}
 so that $\mathcal{P}$ represents the total spin parity of the system. Therefore, we can divide the Hilbert space into sectors of even and odd parity, which are of size $\mathcal{N}_+=\mathcal{N}_-=\mathcal{N}/2$ for half-integer spins,
while
\begin{equation}
\mathcal{N}_\pm=(\mathcal{N}\pm (-1)^N)/2
\label{eq:paritydim}
\end{equation}
for integer spins.

Finally, as $S_n^x$ and $S_n^z$ can always be represented by real matrices, the Hamiltonian displays a time-reversal symmetry $\mathcal{T}H\mathcal{T} \equiv H^*=H$, where $\mathcal{T}$ is antiunitary and fulfills $\mathcal{T}^2=1$. Therefore, the system also possesses a charge-conjugation (particle-hole) symmetry $\mathcal{C}H\mathcal{C}=-H$, where $\mathcal{C}=\mathcal{T}\mathcal{X}$ is an antiunitary operator with $\mathcal{C}^2=1$.

For spins of size $1/2$, it is well known that the rotations defined in  Eq.~\eqref{eqn:genxyz_1} can be written as
$U^{a}(\pi)=i \sigma_a$ with the usual $2\times 2$-dimensional Pauli matrices $\sigma_a$, where one exploits the fact that $\sigma_a^2=\openone$.
In the case of spin 1, where the spin operators are given by Eq.~\eqref{eqn:S1mat_1}, we obtain in contrast
\begin{equation}
	U^{a}(\pi ) = \mathbbm{1} - 2(S^{a})^{2},
\label{eqn:xyz_3}
\end{equation}
where
\begin{equation}
	U^{a}(\pi )U^{b}(\pi ) = U^{c}(\pi ) \quad \forall \ a \neq b \neq c \neq a.
\end{equation}
In other words, the set of these chiral symmetry operators, with the identity operator, is isomorphic to the Klein four-group, which describes the symmetries of a nonsquare rectangle.
We can also exploit the relations $\comm{U^{a}}{S^{a}} = 0$ and $\acomm{U^{a}}{S^{b}}= 0$, provided that $a \neq b$.

\subsection{The zero mode}\label{sec:ZM_1}

A direct consequence of the chiral symmetry is that eigenstates $|\psi_k\rangle$ with energies $E_k$ are paired with eigenstates $|\psi_{\bar k}\rangle=\mathcal{X}|\psi_k\rangle$ with energy $E_{\bar k}=-E_k$. The exception are zero modes with energy $E_k=0$, for which we formally identify the indices $\bar k =k$. Even in the case of degeneracy of these zero modes, they can always be chosen to fulfill
$\mathcal{X}|\psi_k\rangle=\sigma |\psi_k\rangle$, where $\sigma=\pm 1$ distinguishes between two types of zero modes. The number $\nu_\sigma$ of modes of each type is then constrained by the signature of $\mathcal{X}$, according to $\Tr \mathcal{X}=\nu_+-\nu_-$, which serves as a topological index. In particular, in a system with an odd overall Hilbert space dimension, at least one zero mode is always guaranteed to exist, as it is impossible to pair up all states.

In the Ising chain, the Hilbert space dimension $\mathcal{N}$ is even in the case of half-integer spins, and so are the two parity sectors of dimensionality $\mathcal{N}_\pm=(2S+1)^{N-1}$ as soon as $N\geq 2$. In contrast, according to Eq.~\eqref{eq:paritydim}, for integer spins $\mathcal{N}_-$ is even but $\mathcal{N}_+$ is odd. Hence, at least one zero mode, denoted as $|\psi_0\rangle$, is guaranteed to exist in the even-parity sector of chains with integer spins.
In the basis \eqref{eq:basis}, this can be attributed to the existence of the state $|\mathbf{0}\rangle$ (the state where $s_n^z=0$ for all spins), which is the only basis state that is left invariant under the operation with the chiral operator $\mathcal{X}$, which connects all other basis states in pairs with index $\mathbf{s}$ and $\bar{\mathbf{s}} =-\mathbf{s}$.

The zero mode possesses an energy that remains pinned to zero no matter the disorder or interaction strength. This invariance with respect to parameter variations does not occur for any other eigenstate, which leaves the question if this has any bearings on the localization characteristics, in analogy to what is known from single-particle systems.
Therefore, the key question explored in this work is whether the zero mode displays different localization characteristics to the modes with finite energy.

\subsection{Numerical techniques}
We will address the localization properties of the zero mode both by analytical and numerical approaches.
The numerical results are obtained by exact diagonalization from the positive parity sector in chains with an even number  $N$ of spins of size $S=1$, where we apply periodic boundary conditions.
As the effective Hilbert space dimension rises as $(3^N+1)/2$, and only a single zero mode is present in each realization, we obtain disorder averages from chains of limited lengths up to $N=8$, but also show results from individual realizations with $N=10$. For nonzero modes we collect data from  the middle 10\% of the spectrum. Quantities assigned to the zero mode are denoted in the form $Q_0$, while those of nonzero modes are denoted in the form $Q_{\neq 0}$. Disorder averages of any quantity are denoted by an overline, and are obtained from 10000 realizations. Where focusing on individual disorder strengths, we use values $W=1$ for weak disorder (ergodic regime), $W=8$ for moderate disorder, and $W=20$ for strong disorder (localized regime).

\section{Fragmentation of the zero-mode correlations}
\label{sec:fragmentation}

We start with a general key characteristic of the zero mode, which relates to its real-space structure and holds at all strengths of disorder. We first introduce the spin correlation matrix that captures this structure, and discuss its general properties. We then show that the spin correlations of the zero mode fragment into five independent elementary patterns, whilst for nonzero modes there are only two, and verify and illustrate these patterns numerically.

\subsection{Spin correlation matrix}

The real-space spin structure in a given energy eigenstate $|\psi_k\rangle$ is captured by the correlations
\begin{equation}
C_{k,nm} \equiv
\begin{pmatrix}
\expval{S_{n}^{x}S_{m}^{x}} & \expval{S_{n}^{x}S_{m}^{y}} & \expval{S_{n}^{x}S_{m}^{z}} \\
\expval{S_{n}^{y}S_{m}^{x}} & \expval{S_{n}^{y}S_{m}^{y}} & \expval{S_{n}^{y}S_{m}^{z}} \\
\expval{S_{n}^{z}S_{m}^{x}} & \expval{S_{n}^{z}S_{m}^{y}} & \expval{S_{n}^{z}S_{m}^{z}}
\end{pmatrix}
,
\label{eqn:COPDM_1}
\end{equation}
which we consider as blocks of a Hermitian matrix $\mathcal{C}_k$ of dimension $3N\times 3N$. We term the eigenvalues and eigenvectors of this correlation matrix the correlation eigenvalues and eigenvectors, in distinction to the energy eigenvalues and eigenvectors associated with the Hamiltonian.

The following features are useful to note.

(i) According to the relation $(S_n^x)^2+(S_n^y)^2+(S_n^z)^2=2\openone$,
the trace $\Tr \mathcal{C}_k=2N$ is fixed.

(ii) The matrix is well-behaved under local changes $(S_n^x,S_n^y,S_n^z)\to (S_n^x,S_n^y,S_n^z)O_n^T$
of the spin basis by an orthogonal transformation $O_n$ (hence, basis changes that are compatible with the Lie algebra), which transform the correlation matrix  as $C_{k,nm}\to O_n C_{k,nm}O_m^T$. This leaves the eigenvalues of $\mathcal{C}_k$ invariant, while the corresponding  eigenvectors automatically adapt to the chosen local spin orientations.

(iii) Since the eigenstates of the Hamiltonian have a fixed parity, the expectation values $\langle S_n^z S_m^x \rangle=\langle S_n^z S_m^y \rangle=0$. Therefore, the spin correlation matrix decomposes into a direct sum $\mathcal{C}_k=
\Delta_k\oplus Z_k$, given by the block decomposition
$C_{k,nm} = \Delta_{k,nm} \oplus Z_{k,nm}$,
where
\begin{equation}
\begin{aligned}
	& \Delta_{k,nm} =
	\begin{pmatrix}
		\expval{S_{n}^{x}S_{m}^{x}} & \expval{S_{n}^{x}S_{m}^{y}} \\
		\expval{S_{n}^{y}S_{m}^{x}} & \expval{S_{n}^{y}S_{m}^{y}}
	\end{pmatrix},
	\\
	& Z_{k,nm} = \expval{S_{n}^{z}S_{m}^{z}}.
\end{aligned}
\label{eqn:Zdef_1}
\end{equation}

(iv) Utilizing the unitary matrix
\begin{equation}
	V = \frac{1}{\sqrt{2}}
	\begin{pmatrix}
		1 & 1 \\
		-i & i
	\end{pmatrix}
\label{eqn:COB_1},
\end{equation}
we can further introduce the transformed matrix
\begin{equation}
\rho_{k,nm}=V^\dagger \Delta_{k,nm}V=
	\begin{pmatrix}
		\expval{S_{n}^{+}S_{m}^{-}} & \expval{S_{n}^{+}S_{m}^{+}} \\
		\expval{S_{n}^{-}S_{m}^{-}} & \expval{S_{n}^{-}S_{m}^{+}}
	\end{pmatrix}
\end{equation}
with spin-ladder operators $S_n^{\pm}=2^{-1/2}(S_n^x\pm i S_n^{y})$.
Recalling the analogy between spin-ladder operators and fermionic creation and annihilation operators for systems with spin 1/2, this expression resembles a one-particle density matrix (OPDM), equipped with a Bogoliubov-Nambu structure that is appropriate for a system with a nonconserved particle number.

(v) In a canonical basis state $|\mathbf{s}\rangle$, the correlation matrix $\Delta_\mathbf{s}$ is block-diagonal, with each block having a correlation eigenvalue $1$ and an eigenvalue $1-(s^z_n)^2$, so that the correlation eigenvectors are localized on individual spins.
The correlation matrix $Z_\mathbf{s}$ then has elements $Z_{\mathbf{s},nm}=s_n^zs_m^z$, and hence is of rank 1, with a single finite eigenvalue
$Z^{\mathrm{max}}_\mathbf{s}=\sum_n (s_n^z)^2$ (as indicated, we interpret this as the maximal eigenvalue). This counts the number of spins with a nonzero $z$ component.

These features imply that fully localized states are characterized by an approximately quantized correlation spectrum, in close analogy to the OPDM occupation spectrum in a many-body localized system \cite{bera2015,bera2017,hopjan2019}. In contrast, in an ergodic state represented by a random superposition of basis states, the correlation matrix self-averages to $\mathcal{C}_k^\mathrm{erg}\sim (2/3)\openone$. This results in a  correlation spectrum centered around the single value $2/3$, smoothed out by the influence of the residual off-diagonal elements of $\mathcal{C}_k$, which is in close analogy to the smooth OPDM occupation spectrum in an ergodic many-body system.

\subsection{Zero-mode correlations}

As we show next, for the zero mode the correlation matrix $\Delta_0$  further decomposes into four sectors, each pertaining the $S^x$ or $S^y$ component and additionally confined to the sublattice of even or odd sites. This structure follows directly from the symmetry constraints, and hence holds at all strengths of disorder.

To arrive at these features, we first note that for all states time-reversal symmetry implies
\begin{align}
\langle\psi_k|S^y_nS^x_m|\psi_k\rangle=0 \mbox{ if }n\neq m,
\label{eq:trsconstraint}
\end{align}
as this amounts to an expectation value of a Hermitian operator with imaginary matrix elements, evaluated with a real-valued eigenvector.
This constraint does not apply for $n=m$ as the matrix product $S^y_nS^x_n$ is not Hermitian (it is furthermore not simply related to $S^z_n$, in contrast to the case of spin 1/2).
However, for the zero mode  the chiral symmetry  further implies
\begin{align}
\langle\psi_0|S^x_nS^x_m|\psi_0\rangle&=\langle \mathcal{X}\psi_0|S^x_nS^x_m|
\mathcal{X}\psi_0\rangle
\nonumber
\\
&=(-1)^{n-m}\langle\psi_0|S^x_nS^x_m|\psi_0\rangle,
\end{align}
and analogously
\begin{align}
&\langle\psi_0|S^y_nS^y_m|\psi_0\rangle=(-1)^{n-m}\langle\psi_0|S^y_nS^y_m|\psi_0\rangle,
\\
&\langle\psi_0|S^x_nS^y_m|\psi_0\rangle=(-1)^{n-m-1}\langle\psi_0|S^x_nS^y_m|\psi_0\rangle,
\\
&\langle\psi_0|S^y_nS^x_m|\psi_0\rangle=(-1)^{n-m-1}\langle\psi_0|S^y_nS^x_m|\psi_0\rangle,
\end{align}
which are relations that hold for all $n$ and $m$. In combination with the constraint \eqref{eq:trsconstraint}
from time-reversal symmetry, these relations imply that the blocks $\Delta_{0,nm}$ are all diagonal, and furthermore vanish if $n-m$ is odd. Thus, for the zero mode the $\Delta$ correlation matrix decomposes into four independent blocks,
\begin{equation}
\Delta_0=\Delta_0^{x,\mathrm{even}}\oplus\Delta_0^{x,\mathrm{odd}}\oplus\Delta_0^{y,\mathrm{even}}\oplus \Delta_0^{y,\mathrm{odd}},
\end{equation}
where the superscripts denote the supporting spin component and sublattice. Including the spin correlations from $Z_0$,
we can, therefore, identify five independent elementary spin correlation patterns for the zero mode.

\begin{figure*}
  \includegraphics[width=\linewidth]{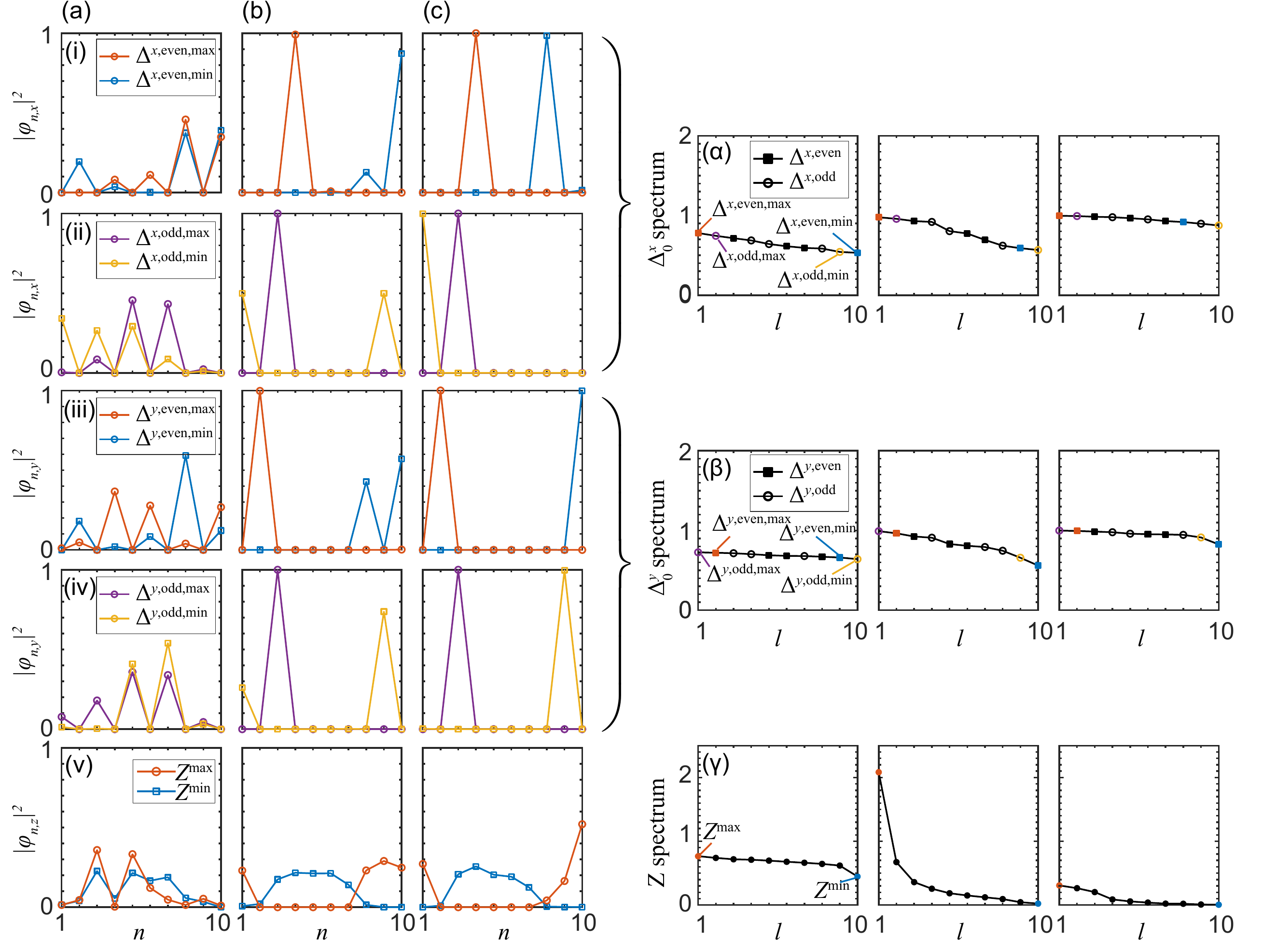}
  \caption{Spin correlations in a zero mode of the spin-1 Ising chain, as quantified by the spin correlation matrix $\mathcal{C}$ in an individual realization of the disorder with strength (a) $W=1$, (b) $W=8$, (c) $W=20$.
  At any disorder, the correlation matrix fragments into 5 sectors.
  Panels (i-v) show the correlation eigenvectors with the largest and smallest correlation eigenvalue in each sector. The corresponding correlation eigenvalue spectra are shown in the adjacent panels ($\alpha$-$\gamma$). For weak disorder, these eigenvalues lie around the ergodic value $2/3$, whilst for strong disorder they approach quantized values 1 (for $\Delta$) and 0 (for $Z$).}
  \label{fig:correlindiv}
\end{figure*}

\begin{figure*}
  \includegraphics[width=\linewidth]{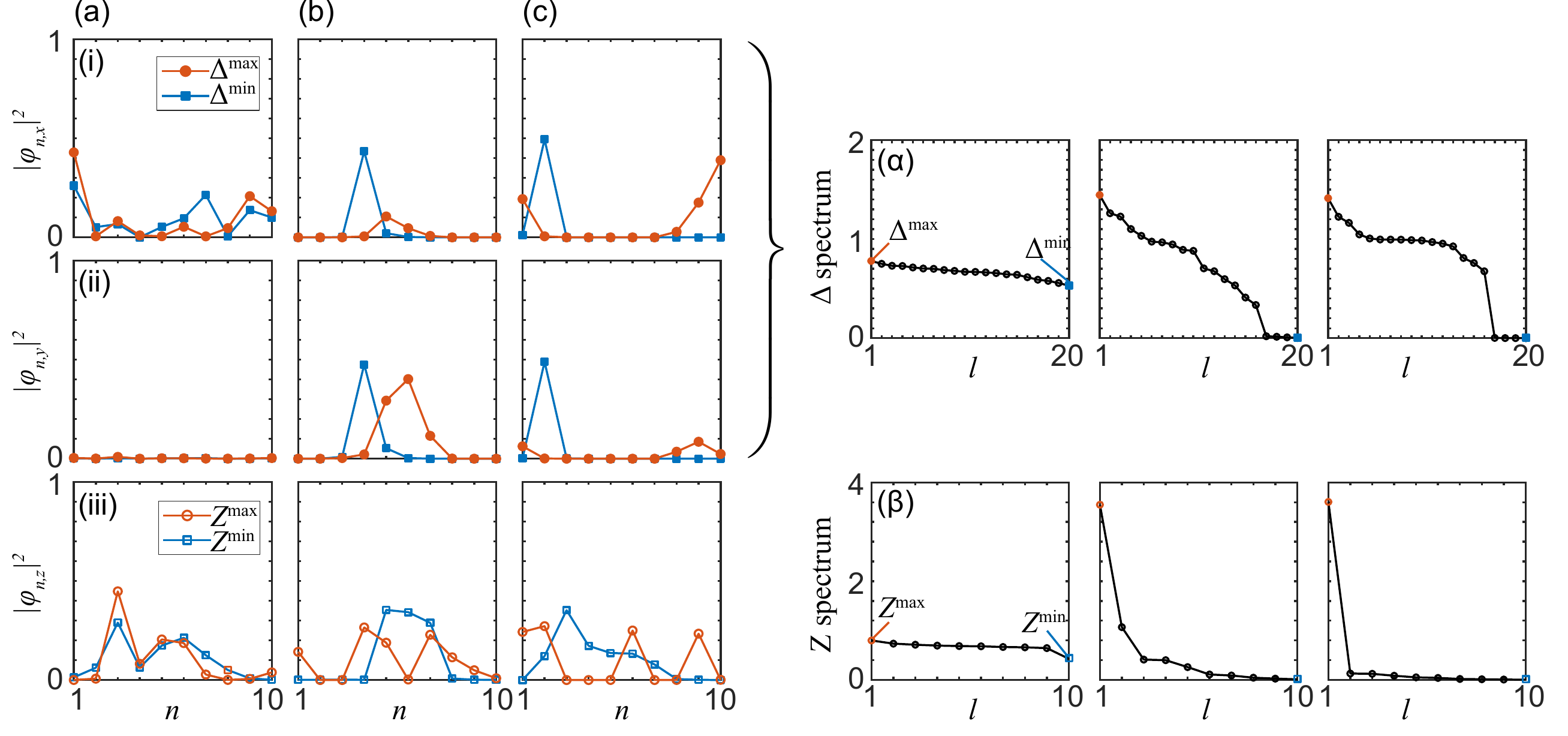}
  \caption{Spin correlations for a nonzero mode close to the band center, in analogy to Fig.~\ref{fig:correlindiv}. Note that the correlation eigenvectors displayed in subpanels (i) and (ii) now belong to the same correlation eigenvalues, hence, represent their $x$ and $y$ components, as these no longer separate, also not with respect to the sublattice. Therefore, only two types of elementary spin correlations exist for such nonzero modes. As shown in subpanels $(\alpha)$ and ($\beta)$, the correlation spectra again become quantized  for strong disorder, reflecting the number of spins with finite $s_z$ in the approached basis state $|\mathbf{s}\rangle$. }
  \label{fig:correlindiv2}
\end{figure*}

\subsection{Numerical illustration}
This structure of the spin correlations is illustrated in Fig. \ref{fig:correlindiv}, where we show correlation eigenstates with minimal and maximal correlation eigenvalues in a typical individual disorder realization at (a) weak, (b) moderate and (c) strong disorder ($W=1,8,20$, respectively).
Subpanels (i-iv) show the four types of correlation eigenvectors from $\Delta_0$, while panel (v) shows correlation eigenvectors from $Z_0$.
The position of these eigenvectors in the occupation spectrum is depicted in the adjacent subpanels ($\alpha$-$\gamma$).

Whilst the predicted fragmented structure holds at all disorder strengths, the correlation eigenvectors from $\Delta_0$ display a noticeable trend from being extended over the whole system for weak disorder, to becoming highly localized on individual spins at strong disorder. In contrast, we notice that the $Z$-correlation eigenvectors more sensitively quantify the hybridization of neighboring spins, a feature that will be important in the subsequent sections.
In conjunction, the correlation spectra from $\Delta_0$ and $Z_0$ both  move away from the ergodic value $2/3$, approaching the quantized values 1 and 0, respectively, as expected for a many-body localized state.

For comparison, Fig. \ref{fig:correlindiv2} shows the analogous spin correlation features in a representative nonzero mode. Note that subpanels (i) and (ii) now refer to the $x$ and $y$ components of the same $\Delta$-correlation eigenvectors, as these correlations no longer separate. Furthermore, each of these eigenvectors now populates both the even and odd sublattices. Otherwise, we notice the same qualitative tendencies as for the zero mode---the $\Delta$ spin correlations again become highly localized for strong disorder, whilst the $Z$  correlations remain more extended, and the corresponding correlation spectra move away from their ergodic values $2/3$ to quantized values, which now depend on the number of finite spins in the approached basis state $|\mathbf{s}\rangle$. In the example, this state has four finite spins, so that there are four nearly-vanishing $\Delta$-correlation eigenvalues, and a dominant $Z$-correlation eigenvalue approaching the value $4$.

By surveying different examples we can certify that these qualitative features are typical for individual states in fixed disorder realizations, with the variations at moderate and strong disorder pointing to different spin hybridization patterns. As indicated above, the $Z_0$-eigenvector  with maximal eigenvalue $Z_0^{\mathrm{max}}$ is particularly useful to characterize the excitation patterns of the zero mode above the reference state  $|\mathbf{0}\rangle$ for the zero mode, and analogously above the reference states  $|\mathbf{s}\rangle$ for nonzero modes. These insights will inform our discussion of the quantitative differences in their localization, on which we focus in the following sections.
\section{Zero-mode delocalization}
\label{sec:delocalization}

We now turn to the second key feature of the zero mode, which pertains to the fact that it is less localized than the nonzero modes. In this section, we establish this feature based on numerical results, while the theoretical explanation is provided in the following section.

\subsection{Measures of localization}

To address this question, we consider a number of complementary indicators of localization, whose general properties we summarize first.

As a general measure of localization, we consider the bipartite von Neumann entanglement entropy \cite{RevModPhys.80.517}. This is defined for each normalized eigenstate $| \psi_k\rangle$ as
\begin{equation}
S_{k} = -\Tr\left(\rho^{(k)}\ln\rho^{(k)}\right),
\label{eqn:VonNeum_1}
\end{equation}
where $\rho^{(k)}=\Tr_{B}\dyad{\psi_{k}}$ is the reduced density matrix of a subsystem $A$, obtained by tracing out the complement $B$.
We take $A$ to be a contiguous subchain of length $N_A=N/2$, hence half the length of the total system.
In delocalized states, the von Neumann entropy is large, and should be well approximated by Page's law for completely ergodic states \cite{Page1993}, $S_k\simeq N_A\ln{3}-\frac{1}{2}$. Therefore, the entropy  grows linearly with the system size, which manifests a volume law.
In contrast, in a localized state the von Neumann entropy is expected to be small, and on average independent of the system size, which manifests an area law. The value of the entropy can then be taken as a proxy for the effective localization length \cite{Szyniszewski2019}. In a basis state $|\mathbf{s}\rangle$, the entanglement entropy $S_\mathbf{s}=0$  vanishes as these states are all separable.

To quantify the degree of Fock-space localization, we make use of the inverse participation ratio (IPR) in the basis \eqref{eq:basis},
\begin{equation}
\mathrm{IPR}_{k} = \sum_{\mathbf{s}}|\langle \mathbf{s}|\psi_{k}\rangle|^{4}.
\label{eqn:IPR_1}
\end{equation}
In the case of perfect Fock-space localization, the IPR goes to unity,
whilst in the case of complete delocalization, the IPR goes to $1/\mathcal{N}$.

We also consider the intensity
\begin{equation}
I_k=|\langle \mathbf{0}| \psi_k\rangle|^2
\label{eq:i0}
\end{equation}
of the states with the special state $|\mathbf{0}\rangle$, which we expect to become large for the zero mode at large disorder, whilst for ergodic states again $I_k\simeq 1/\mathcal{N}$.

\subsection{Numerical results}

\begin{figure}
  \includegraphics[width=\columnwidth]{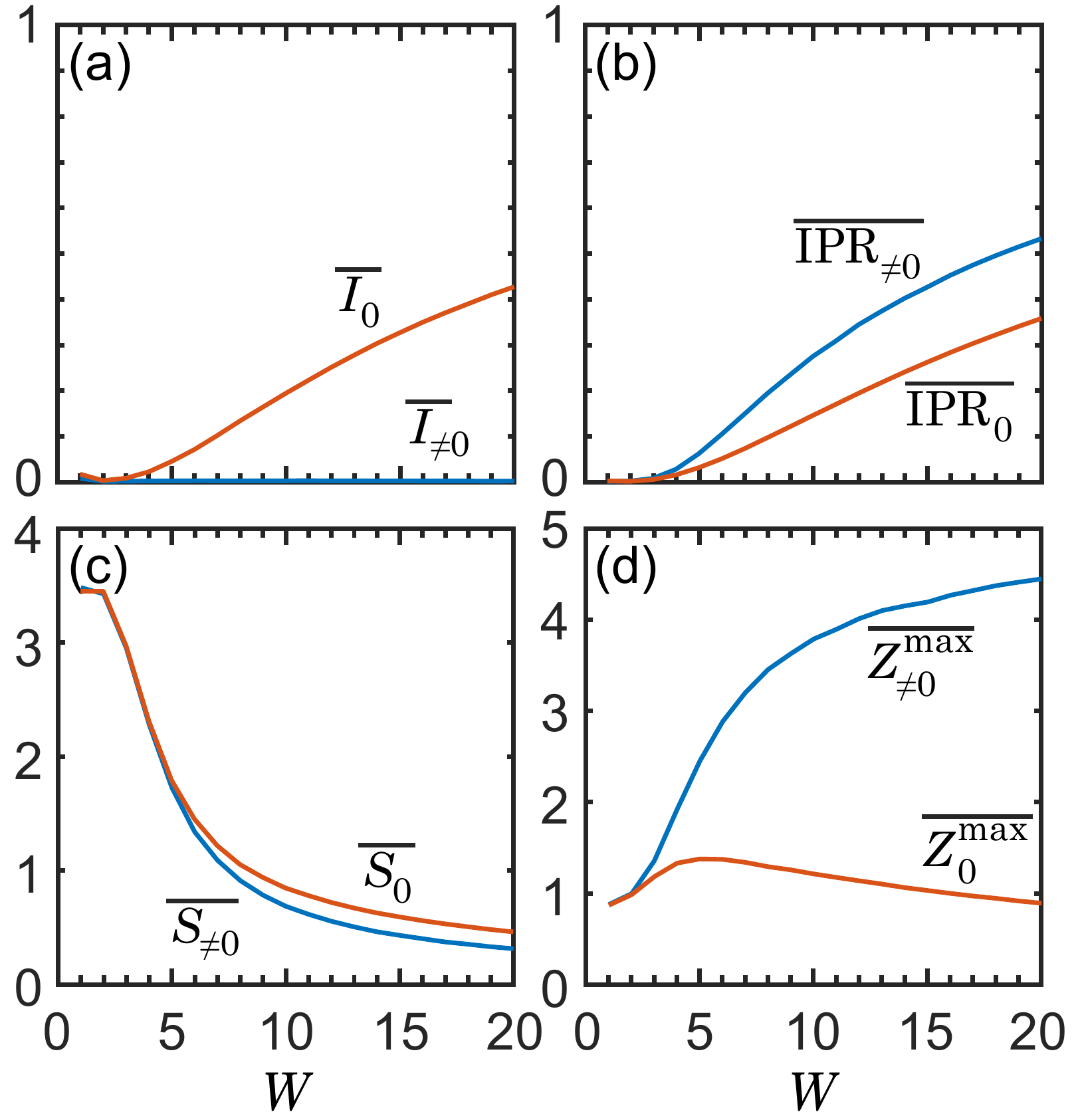}
  \caption{Disorder-averaged measures of localization for the zero mode (light red) and nonzero modes (dark blue) as a function of disorder strength $W$.
   (a)
  Overlap $I_k$ with the basis state $|\mathbf{0}\rangle$, as defined in Eq.~\eqref{eq:i0}.
  For the zero mode, $\overline{I_0}$ increases with increasing disorder strength, reaching $\overline{I_0}\sim 1/2$ at around $W=20$. For nonzero modes, the corresponding average $\overline{I_{\neq 0}}=O(\mathcal{N}^{-1})$ remains negligible at all disorder strengths. As shown in (b), the nonzero modes nonetheless have a larger extent of Fock-space localization,
  as quantified by the inverse participation ratio IPR [see Eq.~\eqref{eqn:IPR_1}], and hence approach an eigenstate $|\mathbf{s}\rangle$ with $\mathbf{s}\neq \mathbf{0}$ more quickly than the zero mode approaches $|\mathbf{0}\rangle$. This relative delocalization of the zero mode is confirmed in panel (c) by the bipartite entanglement entropy, Eq. \eqref{eqn:VonNeum_1}, which is enhanced for the zero mode. Panel (d) shows the maximal $Z$ spin correlation eigenvalue $Z^{\mathrm{max}}_k$, which quantifies the residual hybridization of these states in the strongly localized regime, as further  discussed in Sec.~\ref{sec:pert}.}
    \label{fig:main1}
\end{figure}

\begin{figure*}
  \includegraphics[width=1.8\columnwidth]{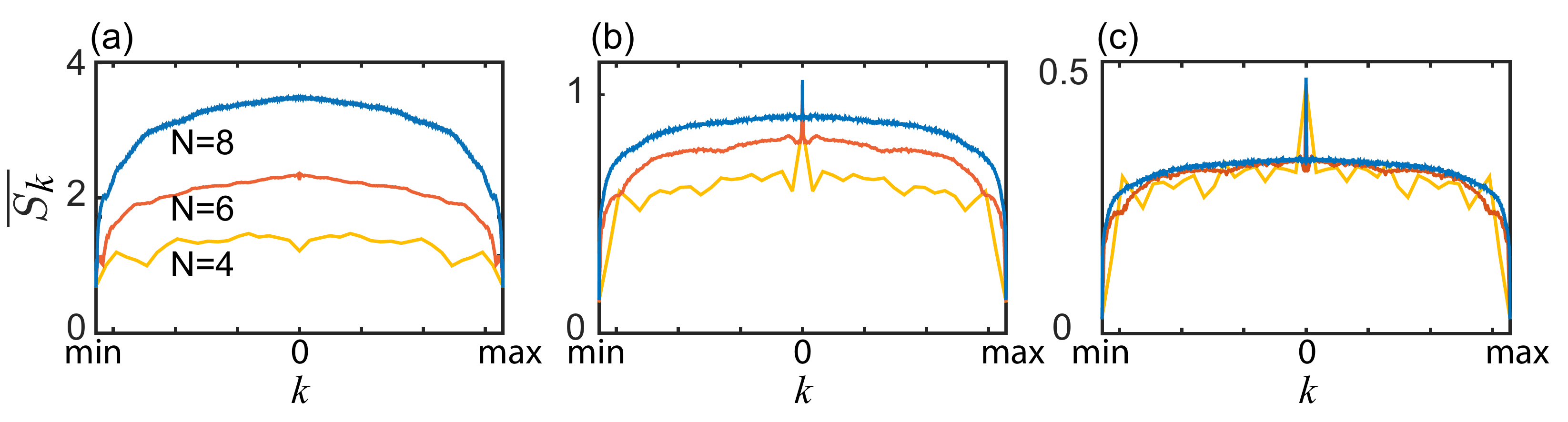}
  \caption{Averaged entanglement entropy of energy eigenstates ordered by their energy,
  with the resulting mode index centered at the zero mode, for disorder strengths
  (a) $W=1$, (b) $W=8$, (c) $W=20$, and system sizes $N=4,6,8$. The entanglement entropy of the
  zero mode is  enhanced in the localized regime, by an amount that is independent of the accessible system sizes.
  }
  \label{fig:main2}
\end{figure*}

\begin{figure*}
  \includegraphics[width=1.8\columnwidth]{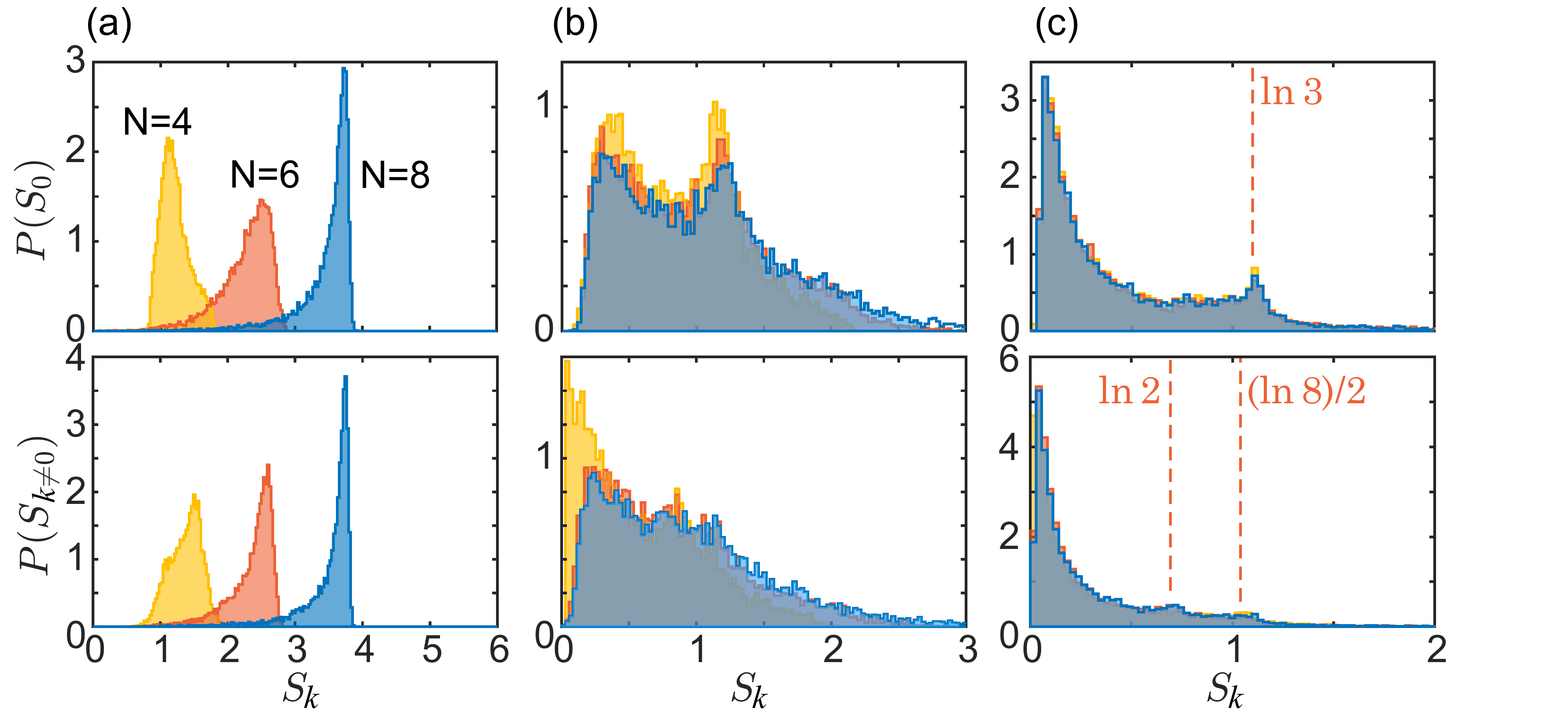}
  \caption{Distributions of the entanglement entropy for zero modes (top panels) and nonzero modes (bottom panel), for parameters as in
  Fig.~\ref{fig:main2}. From moderate disorder, the zero mode shows a significant enhancement of entropies $S\gtrsim 1$,
  with the indicated value $\ln 3$ identified in Sec.~\ref{sec:pert}. Nonzero modes display less pronounced features at
  smaller characteristic values  $\ln 2$ and $(\ln 8)/2$.}
  \label{fig:main3}
\end{figure*}

At strong disorder, the zero mode is expected to have a large overlap
$I_0=|\langle \mathbf{0}| \psi_0\rangle|^2$
with the state $|\mathbf{0}\rangle$, in which the contribution from the field $\mathbf{h}$ vanishes. This is verified in Fig. \ref{fig:main1}(a), which shows that the disorder-averaged $\overline{I_0}$ rises sharply at disorder strengths $W\simeq 4$. In contrast, the corresponding average  $\overline{I_{\neq 0}}\sim O(\mathcal{N}^{-1})$ for the nonzero modes is negligible for all disorder strengths.
Nonetheless, overall the zero mode is noticeably \revision{less localized} in Fock space than the nonzero modes, as evidenced in Fig. \ref{fig:main1}(b) by an inverse participation ratio $\overline{\mathrm{IPR}_{0}}$ that is reduced relative to $\overline{\mathrm{IPR}_{\neq 0}}$, and in Fig. \ref{fig:main1}(c) by a bipartite entanglement entropy $\overline{S_0}$ that is increased relative to  $\overline{S_{\neq 0}}$. Therefore, for strong disorder the nonzero modes approach basis states $|\mathbf{s}\rangle$ with $\mathbf{s}\neq \mathbf{0}$ more quickly than the zero mode approaches the basis state $|\mathbf{0}\rangle$. As we explain in Sec.~\ref{sec:pert}, the residual hybridization of basis states can be quantified by the maximal $Z$ spin-correlation eigenvalue $Z^\mathrm{max}_k$, whose average is shown in Fig. \ref{fig:main1}(d).

In Fig.~\ref{fig:main2}, we show the disorder-averaged entanglement entropy $\overline{S_k}$ as a function of the mode index $k$,
obtained by ordering all states by their energy and centering the resulting index at the zero mode. The entropy of the zero mode
is clearly enhanced in the localized regime, by an amount that is independent of the accessible system sizes\revision{, hence remaining consistent with an area law.} This well-confined
\revision{relative} delocalization peak also confirms
that the enhancement is restricted to exact zero modes, and not shared, e.g., by nonzero modes very close to the band center.

As shown by the statistical distribution functions of the entanglement entropy in Fig. \ref{fig:main3},
this delocalizing tendency can be attributed to an accumulation of zero modes with entropy $S_0$ slightly above 1, to be identified as $S_0\simeq \ln 3$ in the following section. This accumulation is already well pronounced at moderate values of disorder (panel b), and
is well defined at very large values of disorder (panel c),  suggesting that it arises from a specific delocalization mechanism.
In contrast, nonzero modes display accumulations at smaller characteristic values of the entropy, to be identified as $\ln 2$ and $(\ln 8)/2$, which hints towards a competition of several distinct delocalization mechanisms.
We will identify the underlying hybridization patterns in the following section.

\section{Dimer hybridization}
\label{sec:pert}
We explain the relative delocalization of the zero mode based on quasi-degenerate perturbation theory at relatively large disorder.
This reveals a  characteristic dimer hybridization pattern involving three collective basis states localized on neighboring spins, whilst nonzero modes support a much wider range of hybridization patterns.

\subsection{Perturbation theory set-up}

Separating the Hamiltonian into a dominant part $H^{(0)} =\sum_{n=1}^{N}h_{n}S_{n}^{z}$ and a perturbation $V=J\sum_{n=1}^{N}S_{n}^{x}S_{n+1}^{x}$,
the unperturbed eigenstates of the system coincide with the canonical basis states $|\mathbf{s}\rangle$ defined in Eq.~\eqref{eq:basis},
with the zero mode given by $|\psi_0\rangle=|\mathbf{0}\rangle$.
These states carry energy $E^{(0)}_{\mathbf{s}}=\sum h_ns_n^z$, have vanishing entanglement entropy $S_\mathbf{s}^{(0)}=0$, Fock-space localization measures $\mathrm{IPR}^{(0)}_\mathbf{s}=1$ and $I^{(0)}_\mathbf{s}=\delta_{\mathbf{s},\mathbf{0}}$, and quantized correlation eigenvalues from $\Delta$ and $Z$.

These characteristics define the typical features of all modes in the strongly localized regime, where any hybridization is absent. The question is how the modes gradually delocalize due to resonant interactions at weaker disorder. We show that this involves distinct hybridization processes on adjacent spins, leading to characteristic features in the entropy, inverse participation ratio, and spin correlations.

We first identify the resonance conditions in general terms, and then derive the hybridization patterns and their characteristic signatures, which we further support with numerical results.

\subsection{Resonance conditions}

In first-order perturbation theory, the hybridization of the zero mode with other states $|\mathbf{s}\rangle$ is strongly suppressed by energy denominators $E^{(0)}_{\mathbf{s}}$. In principle, hybridization can set in for states with individual $|h_n|\leq J$, for which individual spins can align freely. However, at least two sites need to be involved to retain positive parity, and furthermore these configurations have vanishing perturbation matrix elements unless sites neighbor each other. On the other hand, it should then  suffice that $|h_n|-|h_{n+1}|\simeq J$, instead of both $|h_n|,|h_{n+1}|\simeq J$ individually, implying that such disorder configurations should be dominant as they require fewer constraints.

\begin{figure*}[t]
\includegraphics[width=1.6\columnwidth]{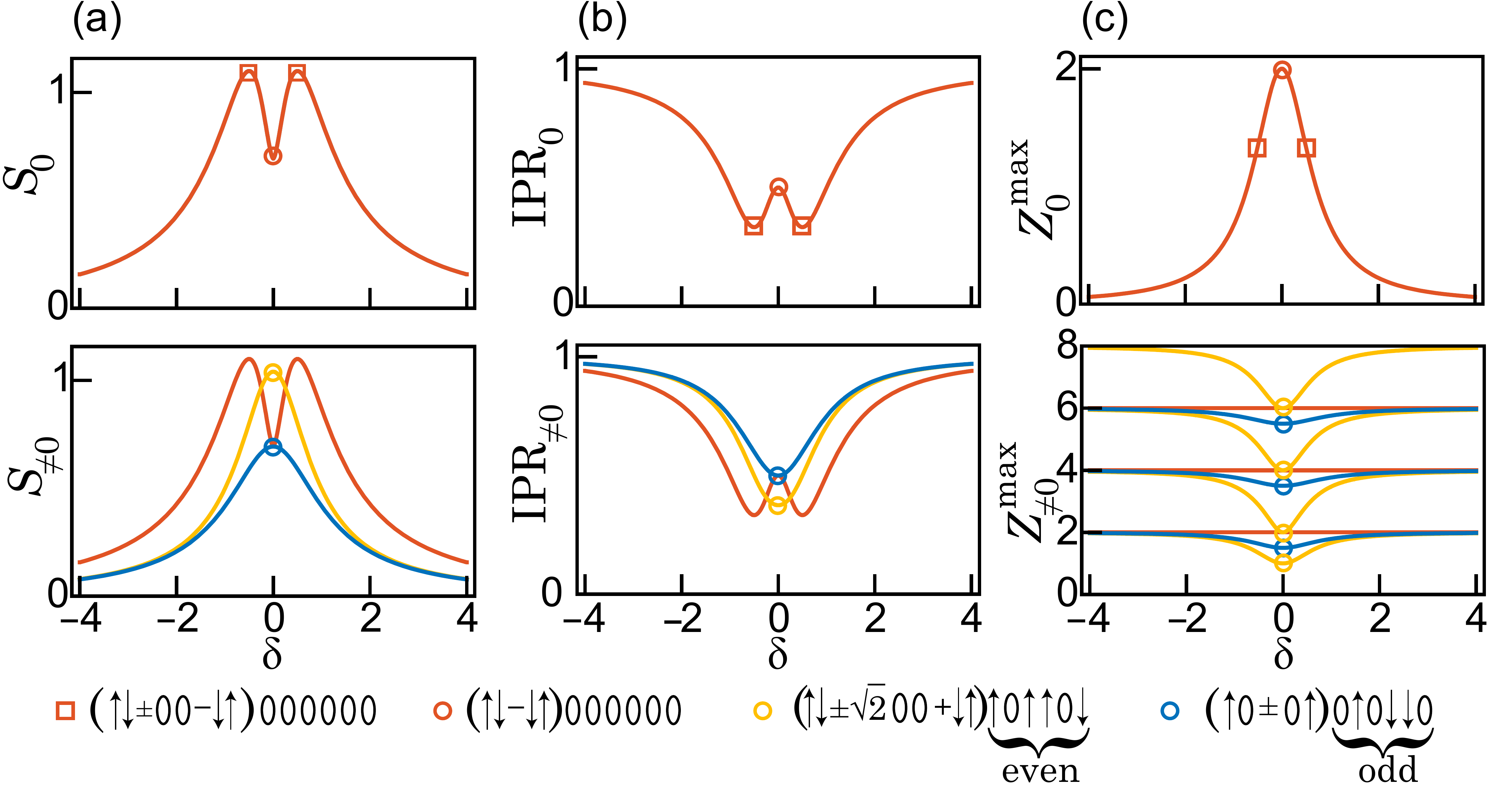}
\caption{Analytical predictions for the localization characteristics of zero modes (top) and nonzero modes (bottom) as a function of the hybridization parameter $\delta$ from quasi-degenerate perturbation theory (see text).
(a) Bipartite entanglement entropy, (b) inverse participation ratio, (c) leading $Z$ spin-correlation eigenvalue $Z_k^\mathrm{max}$.
Quantities arising from the zero-mode hybridizations \eqref{eq:psidimer1} and  \eqref{eq:psidimer2}
are given in medium-dark red.  In the case of nonzero modes, these hybridizations can
still appear when embedded into a state in which an even number of the remaining spins have a finite
$S^z$ component. For the nonzero modes, further configuration scenarios appear from the
hybridizations \eqref{eq:psipmdimer} (again embedded into states with an even number of remaining nonzero spins,
light yellow) and  \eqref{eq:psipmdimer2} (embedded into states with an odd number of remaining nonzero spins, dark blue).
Examples of these hybridization patterns are shown at the bottom of the figure.
}
\label{fig:charanalyt}
\end{figure*}

We can verify the above reasoning by examining all excitations patterns above the background state $|\mathbf{0}\rangle$. Amongst the excitations involving neighboring spins (hence relevant in the first order of the perturbation), only two patterns are allowed by parity, chiral, and time-reversal symmetry, namely those obtained from state $|\mathbf{0}\rangle$ by terms generated via application of the matrix combinations $iS_{n}^x S_{n+1}^y$ and $iS_{n}^y S_{n+1}^x$. These can be conveniently combined into excitation operators
\begin{equation}
\hat\Phi_n^\pm\equiv iS_{n}^x S_{n+1}^y \mp iS_{n}^y S_{n+1}^x,
\end{equation}
leading to the perturbative ansatz
\begin{equation}
|\psi_0\rangle \simeq  \left(1+\sum_n \phi^+_n \hat\Phi^+_n +\sum_n \phi^-_n \hat\Phi^-_n\right)|\mathbf{0}\rangle
,
\end{equation}
where $\phi^\pm_n$ are the amplitudes of the two excitation fields.
Expanding the condition $H|\psi_0\rangle=0$ in orders of the relative interaction strength, we then obtain the perturbatively closed coupled equations
\begin{align}
0&=(\phi^+_n+\phi^-_n)  h_{n+1}+(-\phi^+_n+\phi^-_n)  h_{n}+J,
\\
0&=(\phi^+_n+\phi^-_n)  h_{n}+(-\phi^+_n+\phi^-_n)  h_{n+1},
\end{align}
whereupon
\begin{align}
\phi^+_n=\frac{J}{2(h_n-h_{n+1})},
\label{eq:res1}
\\
\phi^-_n=\frac{-J}{2(h_n+h_{n+1})}.
\label{eq:res2}
\end{align}
Thus, one of the two fields becomes large when $|h_n|\simeq|h_{n+1}|$,
which agrees with the resonance conditions identified above.

\subsection{Zero-mode hybridization patterns}

To describe such resonant disorder configurations more accurately, we resort to quasi-degenerate perturbation theory in the subspace of the hybridizing spins, taken without loss of generality as $(n,n+1)=(1,2)$.
We start with dimer hybridizations of even parity, assuming initially that they are embedded into a chain where the remaining spins are unhybridized,
\begin{equation}
|\psi_0\rangle=|\psi_0\rangle_\mathrm{dimer}\otimes|\mathbf{0}\rangle.
\end{equation}

We first consider the vicinity of the  resonance condition \eqref{eq:res1}, where we
write $h_1=\bar h+\delta/2$, $h_{2}=\bar h-\delta/2$
whilst setting $J=1$.
Ordering the even-parity states as $|1,1\rangle$, $|1,-1\rangle$, $|0,0\rangle$, $|{-1},1\rangle$, $|{-1},-1\rangle$,
the reduced Hamiltonian
\begin{align}
H_{12}^+&=[\bar h(S_1^z+ S_2^z)+\delta/2(S_1^z-S_2^z)+JS_1^xS_2^x]_+
\\
&=
\left(
  \begin{array}{ccccc}
    2\bar h &  0 & 1/2 &  0 &  0 \\
    0 &  \delta &  1/2&  0 & 0\\
    1/2 &  1/2 &  0  & 1/2  & 1/2  \\
    0  & 0  & 1/2  & -\delta  & 0 \\
    0 &  0 & 1/2  & 0 & -2\bar h \\
  \end{array}
\right)
\label{eq:hred1}
\end{align}
then separates into three sectors, with states $|1,1\rangle$ and $|{-1},{-1}\rangle$ gapped out by an energy $\simeq \pm 2 \bar h$,
whilst the zero mode is contained in the quasi-degenerate sector
\begin{align}
\tilde H_{12}^+
&=
\left(
  \begin{array}{ccc}
     \delta &  1/2&  0 \\
      1/2 &  0  & 1/2   \\
    0  & 1/2  & -\delta \\
  \end{array}
\right)
\label{eq:hred2}
\end{align}
spanned by the states $|1,-1\rangle$, $|0,0\rangle$, $|{-1},1\rangle$.
Diagonalizing this sector, we find two states
of finite energy $\pm\sqrt{\delta^2+1/2}$, to which we will come back later, as well as a zero mode
\begin{equation}
|\psi_0\rangle_{\mathrm{dimer}}=|1,-1\rangle-2\delta |0,0\rangle-|{-1},1\rangle
\label{eq:psidimer1}
\end{equation}
of vanishing energy, which we will call the dimer zero mode.
Near the resonance condition \eqref{eq:res2}, the same considerations apply
upon writing $h_1=\bar h+\delta/2$, $h_{2}=-\bar h+\delta/2$
with the roles of the states $(|1,1\rangle,|{-1},-1\rangle)$ and $(|1,-1\rangle,|{-1},1\rangle)$ interchanged,
leading to zero-mode hybridizations
\begin{equation}
|\psi_0\rangle_{\mathrm{dimer}}=|1,1\rangle-2\delta |0,0\rangle-|{-1},{-1}\rangle.
\label{eq:psidimer2}
\end{equation}

In both cases, the bipartite entanglement entropy of the dimer zero mode is given by
\begin{equation}
S_{0,\mathrm{dimer}} =
\ln(2 + 4 \delta^2) -\frac{
 2 \delta^2 \ln (4 \delta^2)}{1 + 2 \delta^2},
 \label{eq:s0dimer}
\end{equation}
and the inverse participation ratio is given by
\begin{equation}
\mathrm{IPR}_{0,\mathrm{dimer}} =\frac{1+8\delta^4}{2(1+2\delta^2)^2}.
 \label{eq:ipr0dimer}
\end{equation}
On the dimer, the $Z$ correlation matrix has a single finite eigenvalue
\begin{equation}
Z_{0,\mathrm{dimer}}^{\mathrm{max}} =\frac{2}{1+2\delta^2},
\label{eq:z0dimer}
\end{equation}
which we interpret as the maximal eigenvalue as for the remaining spins $Z_{0,nn}=0$ vanishes
\footnote{
In this regime this eigenvalue also determines the four-fold degenerate eigenvalue $\Delta_{0,\mathrm{dimer}}=(4-Z_{0,\mathrm{dimer}}^{\mathrm{max}})/4$ of the $\Delta$ correlation matrix (with one eigenvalue per fragmented sector), whilst for the remaining spins $\Delta_{0,nn}=\openone$.}.

The top row in Fig.~\ref{fig:charanalyt} displays these characteristics of
the dimer zero mode as a function of the detuning $\delta$.
The entropy has a stationary point at $\delta=0$ with the value $S_0=\ln 2$,
where $\mathrm{IPR}_0=1/2$ and $Z_{0}^{\mathrm{max}}=2$,
and two stationary points at $\delta=\pm 1/2$ with the value $S=\ln 3$, where $\mathrm{IPR}_0= 1/3$ and $Z_{0}^{\mathrm{max}}=4/3$.

We note that several of these hybridization patterns can be embedded along different positions of the zero mode. The entropy then arrives from the dimers spanning the bipartite partition point, and still adheres to Eq.~\eqref{eq:s0dimer}. The resulting inverse participation ratio is the product of those of all hybridized dimers, so that Eq.~\eqref{eq:ipr0dimer} provides an upper bound for $\mathrm{IPR}_0$. Furthermore, the $Z$ correlation matrix decomposes into independent blocks, so that Eq.~\eqref{eq:z0dimer} provides a lower bound for the maximal $Z$ correlation eigenvalue $Z_0^{\mathrm{max}}$.

\begin{figure}[t]
\includegraphics[width=\columnwidth]{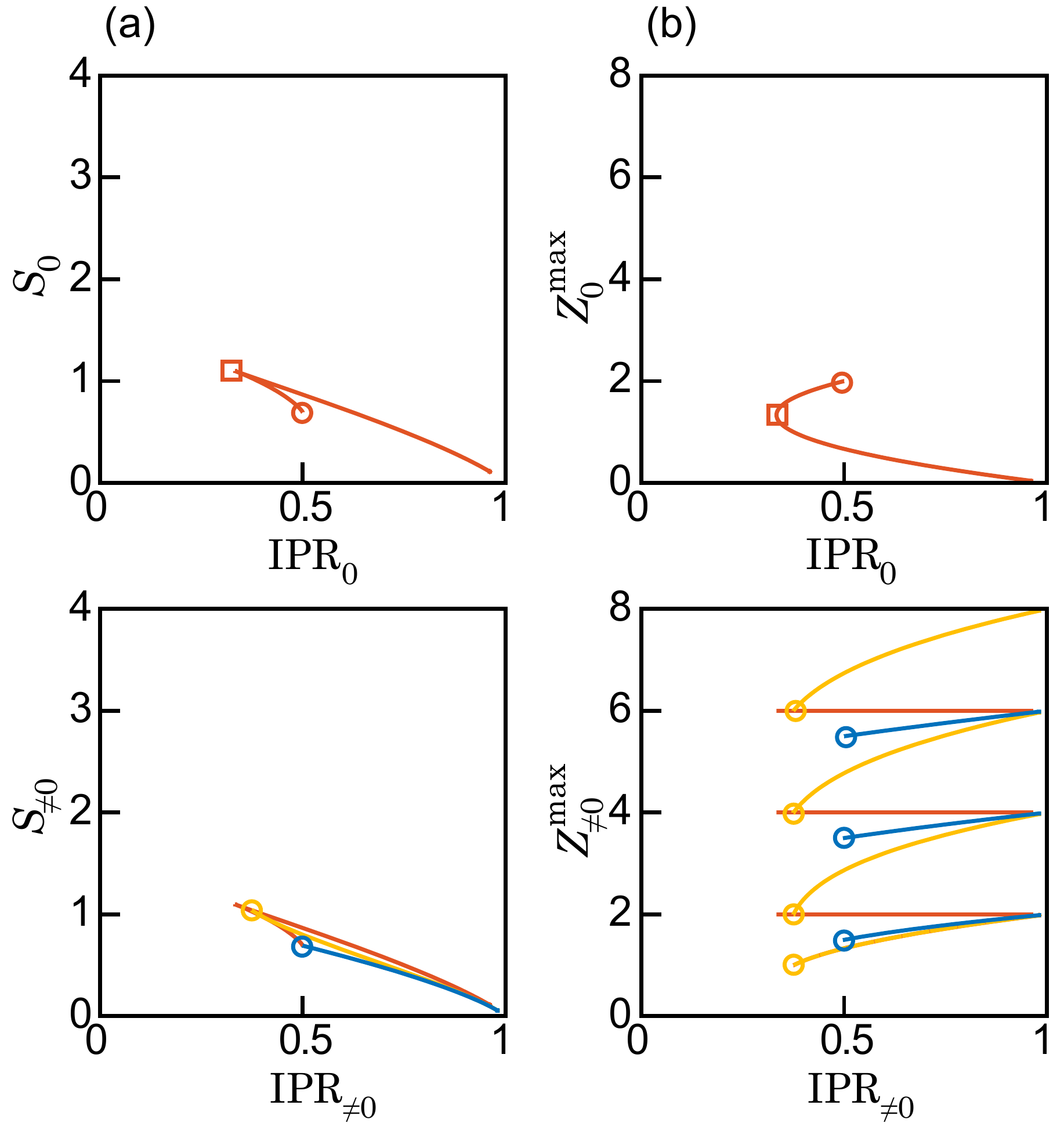}
\caption{Analytical predictions for correlations between the localization characteristics of zero modes (top) and nonzero modes (bottom), following from the results shown in Fig.~\ref{fig:charanalyt}.}
\label{fig:correlanalyt}
\end{figure}

\subsection{Hybridization patterns of nonzero modes}

We next identify the dominant hybridization patterns of nonzero modes,
\begin{equation}
|\psi\rangle=|\psi\rangle_\mathrm{dimer}\otimes|\mathbf{s}'\rangle,
\end{equation}
of which there is a much wider variety, each having its own characteristic signatures.

We start with dimer hybridizations of even parity, embedded into a chain where the remaining spins $|\mathbf{s}'\rangle$ also have even parity.
Assuming again first $h_1=\bar h+\delta/2$, $h_{2}=\bar h-\delta/2$, the dimers of even parity are still described by the reduced Hamiltonian \eqref{eq:hred1}, but all five resulting dimer states have to be taken into account.
Alongside the hybridization pattern \eqref{eq:psidimer1}, this includes the gapped states $|1,1\rangle$ and  $|{-1},{-1}\rangle$, which remain separable,
as well as the two finite-energy  hybridizations
\begin{align}
&|\psi_{+,\pm}\rangle_\mathrm{dimer}=
\\
& (\delta\pm \sqrt{\delta ^2+\frac{1}{2}})|1,{-1}\rangle+ |00\rangle +
 (-\delta\pm \sqrt{\delta ^2+\frac{1}{2}})|{-1},1\rangle
 \nonumber
\label{eq:psipmdimer}
 \end{align}
from the sector \eqref{eq:hred1}.
In the dimer subspace $|1,0\rangle$, $|0,1\rangle$, $|0,{-1}\rangle$, $|{-1},0\rangle$ with odd parity, the reduced Hamiltonian takes the form
\begin{align}
H_{12}^-=[\bar h(S_1^z+ S_2^z)+\delta(S_1^z-S_2^z)+JS_1^xS_2^x]_-
\\
=
\left(
  \begin{array}{cccc}
    \bar h+\delta/2 & 1/2  & 1/2  & 0 \\
    1/2  & \bar h-\delta/2  & 0  & 1/2 \\
    1/2  & 0 &  -\bar h+\delta/2  & 1/2\\
    0  & 1/2 & 1/2  & -\bar h-\delta/2 \\
  \end{array}
\right),
\end{align}
leading to pairwise hybridization
\begin{align}
&|\psi_{-,\pm,1} \rangle_\mathrm{dimer}=(\delta\pm\sqrt{1+\delta^2})|1,0\rangle+|0,1\rangle,
\\
&|\psi_{-,\pm,2} \rangle_\mathrm{dimer}=(\delta\pm\sqrt{1+\delta^2})|0,{-1}\rangle+|{-1},0\rangle
\label{eq:psipmdimer2}
\end{align}
only.

\begin{figure}[t]
\includegraphics[width=\columnwidth]{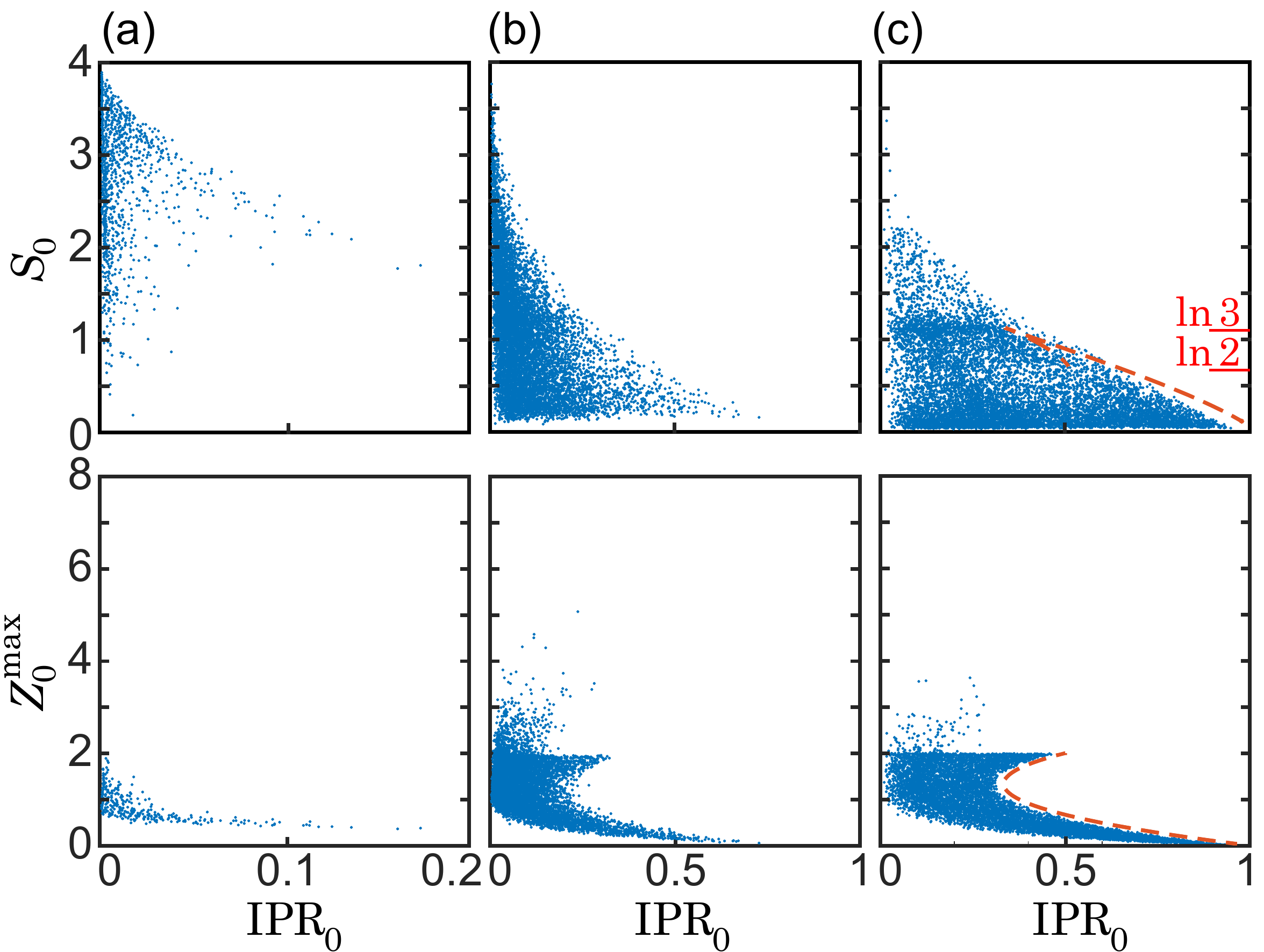}
\caption{Scatter plot of localization characteristics of the zero mode for disorder strengths (a) $W=1$, (b) $W=8$, (c) $W=20$.
In (c), the dashed lines indicate the predicted analytical bounds, see the top panels in Fig.~\ref{fig:correlanalyt}(a,b).
Note the different scale on the horizontal axis in panel (a).
}
\label{fig:correlz}
\end{figure}

\begin{figure}[t]
\includegraphics[width=\columnwidth]{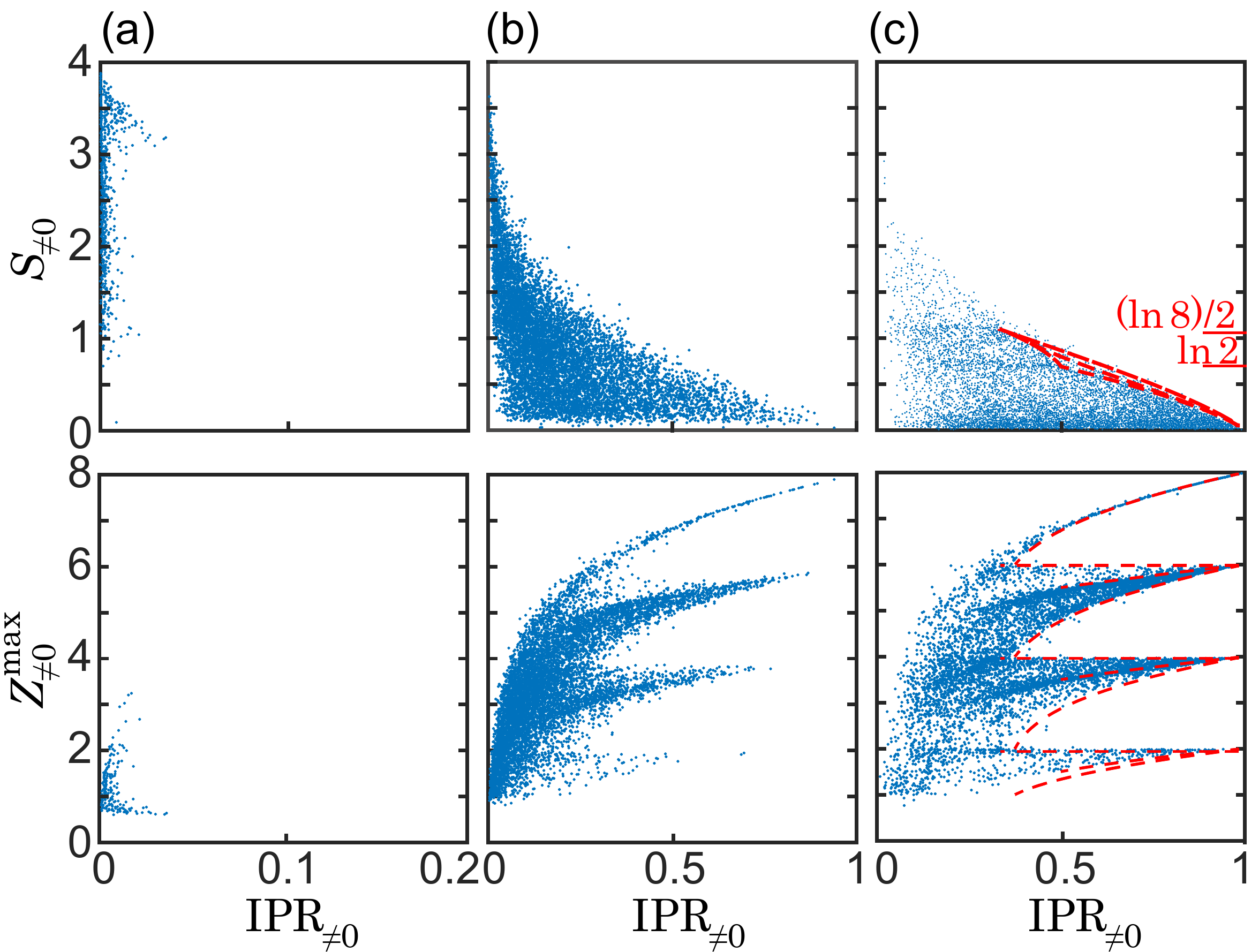}
\caption{Scatter plot of localization characteristics of nonzero modes in analogy to Fig.~\ref{fig:correlz}, with the analytical
bounds in (c) taken from the bottom panels of Fig.~\ref{fig:correlanalyt}(a,b).}
\label{fig:correlnz}
\end{figure}

Overall, we therefore arrive at seven hybridization patterns and two nonhybridized states, reflecting the full dimensionality of the dimer subspace.
For the second resonance case $h_1=\bar h+\delta/2$, $h_{2}=-\bar h+\delta/2$, the same considerations apply upon interchanging dimer basis states $|s,s'\rangle\leftrightarrow |s,-s'\rangle$.

The characteristic features of these finite-energy hybridization patterns are shown in the bottom row of Fig.~\ref{fig:charanalyt}.
Hybridizations based on $|\psi_0\rangle_\mathrm{dimer}$ still produce the same entropies and inverse participation ratios as for the zero mode,
while the largest eigenvalue of the $Z$ correlation matrix now arises from the remainder of the chain, where it counts the number of finite spins, thus giving rise to the straight lines at even integers. For the hybridizations $|\psi_{+,\pm}\rangle_\mathrm{dimer}$ of even parity, the entropy is stationary around $\delta=0$, where $S_k=(\ln 8)/2$ whilst the inverse participation ratio takes the value $\mathrm{IPR}_k=3/8$.
For the hybridizations $|\psi_{-,\pm,k}\rangle_\mathrm{dimer}$ of odd parity, a similar behavior is observed with stationary entropies $S_k=\ln 2$ and inverse participation ratios $\mathrm{IPR}_k=1/2$. In both these hybridization patterns, the eigenvalue $Z_k^\mathrm{max}$ depends both on the hybridization strength $\delta$ and the number of finite spins in the remainder of the chain.
Considering that several hybridized dimers can occur along the chain, we again interpret the predicted inverse participation ratios and $Z$ correlation eigenvalues as upper bounds and lower bounds, respectively.

\subsection{Summary and numerical verification}

Summarizing the results from this section, we arrive at the following detailed predictions.

For the zero mode, delocalization occurs via dimer hybridization patterns with typical entropies $S_0\sim\ln 3$ or $S_0\sim\ln 2$, as already observed numerically in the upper panels of Fig.~\ref{fig:main3}. Entropies $S_0\sim\ln 3$ are further expected to correlate with inverse participation ratios bounded as $\mathrm{IPR}_0\lesssim 1/3$ and leading $Z$ correlation eigenvalues bounded by
$Z_{0}^\mathrm{max} \gtrsim 4/3$, while for entropies $S_0\sim\ln 2$ we expect $\mathrm{IPR}_0\lesssim 2$ and $Z_{0}^\mathrm{max} \sim 2$. More generally, these quantities should be correlated as shown in the upper panels of Fig.~\ref{fig:correlanalyt}.
These predictions are verified in Fig.~\ref{fig:correlz},
where we show scatter plots of the described quantities from $10^4$ realizations for chains of length $N=8$. The expected correlations are already well established for moderate values of disorder $W=8$.

Furthermore, nonzero modes should predominantly display entropies around $S_k=\ln 2$, which can be achieved by the widest variety of hybridization patterns, followed by $S_k=(\ln 8)/2$, whilst $S_k=\ln 3$ should occur relatively less frequently, as indeed observed in the lower panels of Fig.~\ref{fig:main3}.
The expected correlations with $\mathrm{IPR}_k$ and $Z_{k}^\mathrm{max}$ are depicted in the lower panels of Fig.~\ref{fig:correlanalyt}.
These predictions are verified in Fig.~\ref{fig:correlnz}.

Comparing the results in Figs.~\ref{fig:correlz} and \ref{fig:correlnz}, we find that the
zero modes and nonzero modes are most clearly discriminated by their distinct correlations between the inverse participation ratio $\mathrm{IPR}_k$ and the leading $Z$-correlation eigenvalue $Z_k^\mathrm{max}$.

Having confirmed these key predictions, we return to Fig.~\ref{fig:charanalyt} to observe that the dominant hybridization patterns of the zero modes are  appreciable over a larger range of detunings $\delta$ than for the nonzero mode. This verifies that the zero mode hybridizes more readily than the nonzero modes, and then exhibits more delocalized Fock-space configurations, which provides the general explanation for the numerical observation of this effect in the previous section.

\section{Discussion and conclusions}
\label{sec:conclusions}

In summary,
in many-body systems zero modes protected by a chiral symmetry can localize, but then do so with distinctively different characteristics than nonzero modes. In particular, the zero modes are more delocalized both in terms of their real-space and Fock-space signatures. We explained these differences by the characteristic symmetry-restricted mechanisms allowing the localized basis states to hybridize. These symmetry constraints can be extended to all disorder strengths by considering the fragmentation of real-space correlations.

We developed and demonstrated these effects for the example of a disordered spin-1 Ising chain. For spin 1/2-chains, the chiral symmetry is already present, but symmetry-protected zero modes do not occur as the Hilbert  space dimension is even in both parity sectors. In spin-1/2 chains, the nonzero modes are known to delocalize by a single dominant hybridization pattern, involving dimers with bipartite entanglement entropy $S_k=\ln 2$ \cite{Laflorencie2005}. In contrast, in the spin-1 chain, the delocalization \revision{mechanism} of zero modes involves a dominant hybridization pattern with entropy $S_0=\ln 3$, whilst
nonzero modes involve a competition of various hybridization patterns, including such with entropy $S_k=(\ln 8)/2$. Even though the underlying hybridizations differ, these entanglement values are reminiscent to those encountered in the fragmented ground state of the spin-1 system by
Affleck, Kennedy, Lieb, and Tasaki (AKLT model) \cite{PhysRevLett.59.799}, as well as for other spin-1 systems, where such entanglement entropy values can be found between a single spin and the remainder of the system \cite{PhysRevLett.93.227203,Li2018}.
Furthermore, for the studied system the fragmentation of real-space correlations occurs both with respect to the spin orientation as well as with respect to the even and odd sublattices, where the latter is particularly noteworthy as statistically the system is translationally invariant.

A fundamental tenet for disordered interacting quantum systems is the expectation that many-body states close in energy share the same statistical signatures. The symmetry-protected zero-modes discussed here provide a mechanism to equip individual states with their own characteristic signatures. It would be interesting to explore whether the remarkable differences between zero modes and nonzero modes become further accentuated for larger integer spins, and whether these observations also extend to appropriately designed itinerant fermionic systems.

\begin{acknowledgments}
We gratefully acknowledge discussions with Jens Bardarson.
This research was funded by UK Engineering and
Physical Sciences Research Council (EPSRC) via Grants
No. EP/P010180/1 and EP/L01548X/1. Computer time was provided by
Lancaster University's High-End Computing facility.
\end{acknowledgments}


\appendix

\begin{figure}[t]
  \includegraphics[width=0.8\linewidth]{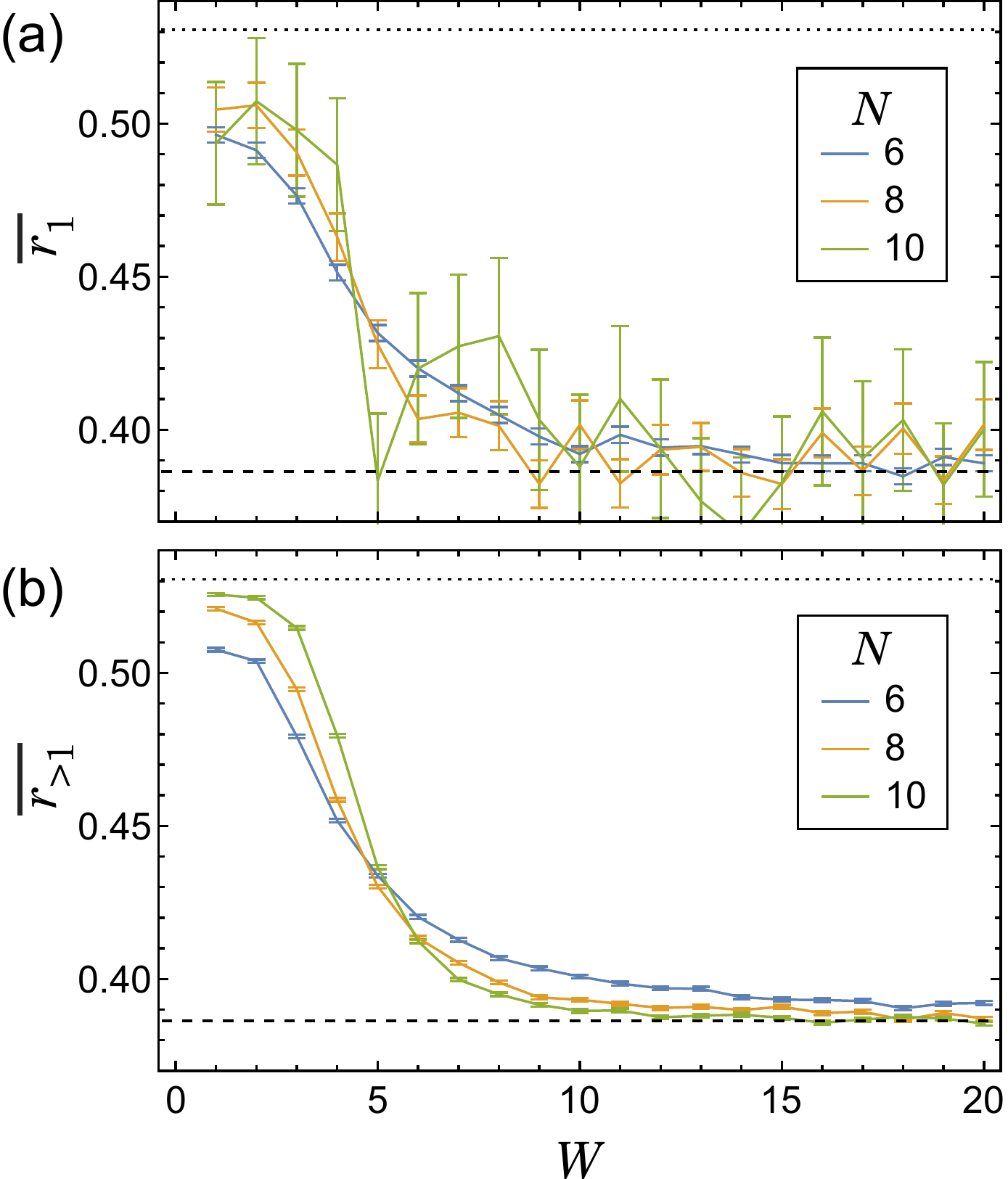}
  \caption{\revision{Disorder-averaged level-spacing ratio \eqref{eq:raverage} as a function of disorder strength $W$, for (a) $r_1$ (including the spacing of the zero mode to its next neighbour) and (b) $r_k$ with $k>1$ (only involving spacing between nonzero modes, which are furthermore restricted to the middle 10\% of the spectrum). The dotted line indicates the value expected for an ergodic system modelled by the Gaussian orthogonal ensemble of random matrix theory, while the dashed line indicates the value for a localized system with Poissonian level statistics. For the three system sizes $N=6,8,10$, the data shown is obtained from $\sim 10^4,10^3,10^2$ realizations, respectively.}
}
\label{fig:raverage}
\end{figure}

\begin{figure}[t]
  \includegraphics[width=\linewidth]{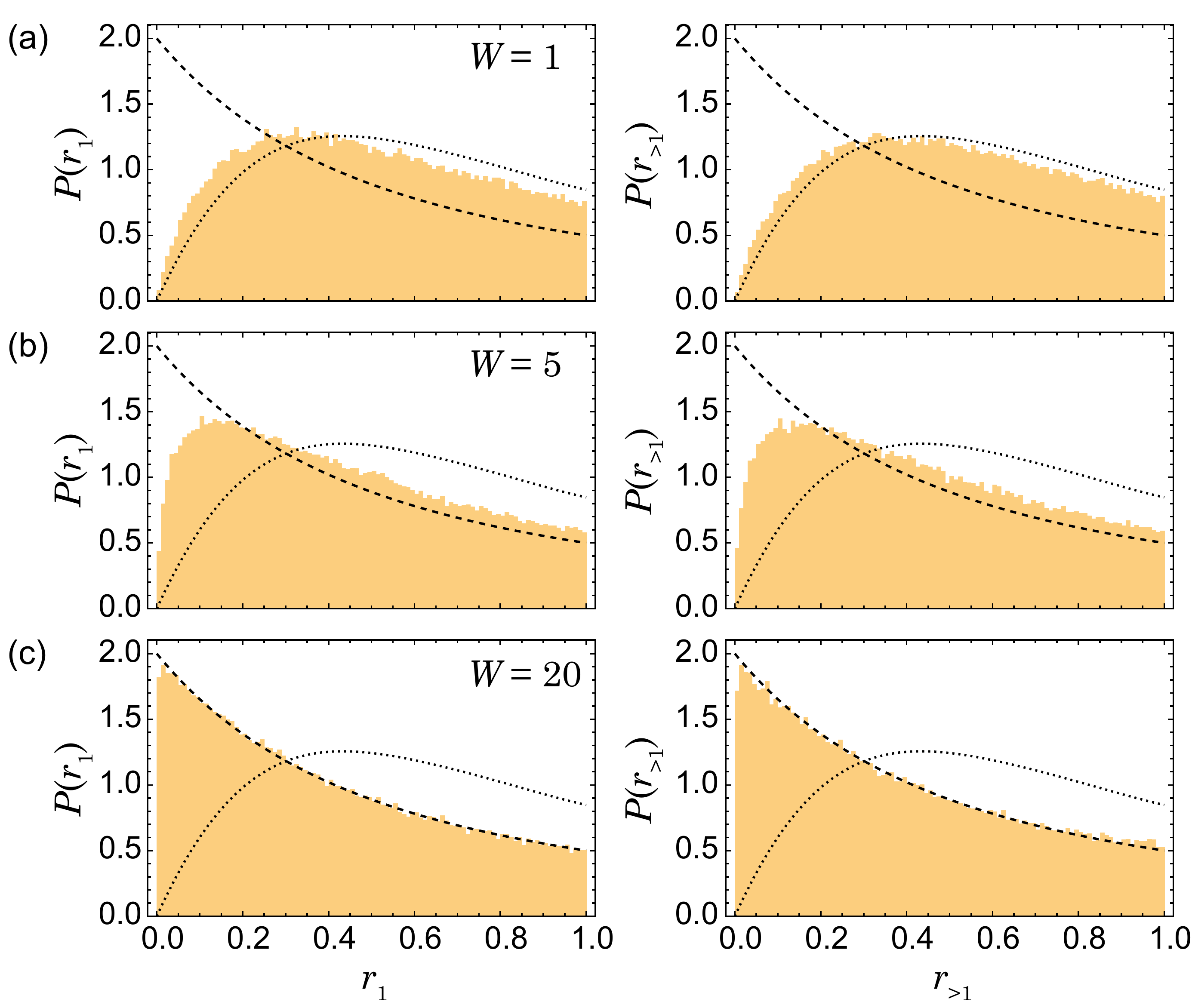}
  \caption{
\revision{Distribution of level-spacing ratios corresponding to Fig.~\ref{fig:raverage}, with the system size fixed at $N=6$ and the disorder strength set to (a) $W=1$, (b) $W=5$, and (c) $W=20$, using $2\times 10^5$ realizations. The curves indicate the expected distribution for an ergodic system modelled by the Gaussian orthogonal ensemble of random matrix theory (dotted), as well as a localized system with Poissonian level statistics (dashed).}
}
\label{fig:rdist}
\end{figure}

\revision{
\section{Level statistics}
The results in this work firmly indicate that all states in the spin-1 Ising chain \eqref{eqn:Ising_1}
become many-body localized when the disorder becomes sufficiently strong, irrespective of whether they are zero modes or nonzero modes (see, e.g., Fig.~\ref{fig:main1}). As this specific model has not been considered before, we here provide further supporting evidence based on the statistics of the standard level-spacing ratio $r_k$, defined as \cite{Oganesyan2007,Pal2010,Luitz2015}
\begin{equation}
r_k=\min\left(\frac{E_{k+1}-E_{k}}{E_{k}-E_{k-1}},\frac{E_{k}-E_{k-1}}{E_{k+1}-E_{k}}\right)
,
\label{eq:raverage}
\end{equation}
where the energies $E_k$ are ordered by magnitude.
In an ergodic system, the averaged ratio is expected to be large due to level repulsion, with $\overline{r_k}\approx  0.5307(1)$ if modelled via the Gaussian orthogonal ensemble, whilst in a many-body localized system it is expected to drop to a smaller value, approaching $\overline{r_k}=2 \ln 2 - 1 \approx 0.38629$ corresponding to Poissonian level statistics \cite{Atas2013}.

We restrict our attention to the parity sector including the zero mode.
Given the chiral symmetry, we arrange the indices $k$ so that $E_0=0$ denotes the zero mode and $E_k=-E_{-k}$ denotes the levels paired by the spectral symmetry. Due to this pairing, $r_0\equiv 1$ in each realization, so we instead resort to $r_1$ to characterize the zero mode (which is involved via the spacing $E_1-E_0$). We contrast this with the statistics of $r_k$ with $k\geq 1$ for the nonzero modes, which we constrain to the middle 10\% of the spectrum
 (note that the chiral symmetry furthermore implies $r_k=r_{-k}$).

 In Fig.~\ref{fig:raverage}, the disorder-averaged spacing ratios are shown as a function of disorder strength for three system sizes $N=6,8,10$. The statistical fluctuations for $r_1$ are large as only a single value is obtained for each realization. Nonetheless, both figures consistently point towards states becoming localized at about the same strength of disorder, with the averaged ratios of different system size crossing near a point  of inflection at around $W\simeq 5$.

In Fig.~\ref{fig:rdist}, we show the full statistical distribution of the spacing ratios for the smallest system size $N=6$, where enough data can be collected, for representative values of the disorder strength $W=1,5,20$. The results for $r_1$ and $r_{>1}$ resemble each other closely in all three cases, being consistent with ergodic behavior for $W=1$ as well as many-body localized behavior for $W=20$, and displaying similar intermediate statistics for $W=5$.
}

\clearpage

%

%

\end{document}